\documentclass[preprint2]{aastex}
\usepackage{natbib}
\shorttitle{Radial Stellar Pulsation and Three-dimensional Convection. II.}
\shortauthors{Geroux and Deupree}

\begin{document}

\title{RADIAL STELLAR PULSATION AND THREE-DIMENSIONAL CONVECTION. II. TWO-DIMENSIONAL CONVECTION IN FULL AMPLITUDE RADIAL PULSATION}

\author{Chris M. Geroux\altaffilmark{1} and Robert G. Deupree}
\affil{Institute for Computational Astrophysics and Department of Astronomy and Physics, Saint Mary’s University, Halifax, NS B3H 3C3 Canada}
\email{geroux@astro.ex.ac.uk}
\altaffiltext{1}{Now at Physics and Astronomy, University of Exeter, Stocker Road, Exeter, EX4 4QL, UK }

\begin{abstract}
We have developed a three-dimensional radiation hydrodynamics code to simulate the interaction of convection and radial pulsation in classical variable stars. One key goal is the ability to carry these simulations to full amplitude in order to compare them with observed light curves. Previous multi-dimensional calculations were prevented from reaching full amplitude because of drift in the radial coordinate system, due to the algorithm defining radial movement of the coordinate system during the pulsation cycle. We have removed this difficulty by defining our radial coordinate flow algorithm to require that the mass in a spherical shell remains constant for every time-step throughout the pulsation cycle. We have used our new code to perform two-dimensional (2D) simulations of the interaction of radial pulsation and convection. We have made comparisons between light curves from our 2D convective simulations with observed light curves and find that our 2D simulated light curves are better able to match the observed light curve shape near the red edge of the RR Lyrae instability strip than light curves from previous one-dimensional time dependent convective models.
\end{abstract}


\section{INTRODUCTION}
Classical variables stars such as Delta Cepheids and RR Lyraes have long played an important role in astronomy as distance indicators on galactic and extragalactic scales for Population I and Population II stars, respectively. RR Lyrae provide an independent test of the classical Cepheid distance scales for nearby galaxies \citep[e.g. Magellanic Clouds and M31][]{Pritchet-1987,Storm-2006}, and are also used to calibrate secondary distance indicators such as globular cluster luminosity functions \citep{Criscienzo-2006}. Combining the location of the main-sequence turnoff in the HR diagrams of globular clusters with evolutionary models provides estimates of ages of globular clusters and puts constraints on stellar ages. The age estimates of globular clusters, which are impacted by accurate distances, place constraints on cosmological models.

Thus, understanding variable stars, and in particular RR Lyrae variables, is important for many aspects of astronomy and for constraining some internal characteristics of stars such as the role of convection. To gain a full understanding of variable stars, models must be compared with observations. Much work toward comparing models, nearly all one-dimensional (1D), and observations has been done and considerable progress has been made. However, there are some issues remaining. In particular the current 1D models for the interaction between time dependent convection and radial pulsation have difficulties reproducing observed properties for cool RR Lyrae variable stars \citep{Marconi-2009}. The current work explores the interaction of convection and radial pulsation by directly simulating the convection in a manner similar to that used to hydrodynamically study solar convection \citep[e.g.,][]{Stein-1998,Nordlund-2009}. We will make comparisons of our full amplitude, multi-dimensional, convective RR Lyrae models with observations of RR Lyrae variables in M3. In the next Section we introduce the observations of RR Lyrae stars to which we will compare our models.

\subsection{Observations of RR Lyrae Stars}
\label{sec:obs}
The extensive observational efforts on RR Lyrae stars has shown that they have relatively constant mean magnitudes, and thus luminosities across the instability strip; typical values for the luminosity of RR Lyrae stars are from 40  to 70$L_\odot$ \citep[e.g.][for the RR Lyrae variables in M3]{Cacciari-2005}. \citeauthor{Cacciari-2005} find an average mass for these stars of 0.71$\pm$ 0.03$M_\odot$, however they mention that there is considerable uncertainty associated with the masses of RR Lyrae stars and that determining masses from double mode pulsators favor masses on the order of $0.74M_\odot$, while determining masses from evolution favor masses around $0.68M_\odot$.

There are uncertainties in measuring magnitudes and colors in RR Lyrae, as for all stars, from instrumental limitations, reddening, and atmospheric effects, for example. For variable stars in particular there are additional uncertainties with conversion of colors to effective temperatures of the equivalent static star. Observationally measured color indices are transformed using color-effective temperature relations, which are provided by grids of static atmosphere models. Thus, how the color-effective temperature relations for pulsating stars relate to those of static models is an issue. The color of an RR Lyrae variable changes through out the pulsation cycle; at which phase should the color be measured or using what sort of average? The usual procedure is to compute a mean color index. This can be done by either computing the averages with the magnitudes themselves, the intensities, or the differences between magnitudes or intensities. \cite{Sandage-1990} and \cite{Bono-1995} have both explored different methods for obtaining mean colors to represent those of the static star. While \citeauthor{Sandage-1990} approached this empirically, and \citeauthor{Bono-1995} theoretically, both find similar methods to be best. Intensity averages of $B$ and $V$ separately plus some sort of amplitude-related correction reproduce the equivalent static colors reasonably well. We will be comparing mostly with observations by \cite{Cacciari-2005} of M3 RR Lyrae variables, who use the method developed by \cite{Bono-1995}.

Knowing the observational location of the RR Lyrae instability strip is important for validation of theoretical models. Moving from the cool edge to the hot edge of the instability strip we first have the fundamental red edge (FRE), followed by the first overtone red edge (FORE), the fundamental blue edge (FBE), and then finally the first overtone blue edge (FOBE). \cite{Sandage-1990} studied the globular clusters M3 and M15, deriving effective temperatures from observed colors for a number of RR Lyrae variables in the fundamental and first overtone modes and for non-variable stars in the horizontal branch. From \citeauthor{Sandage-1990}'s data we deduce the locations of the edges for M3 with the FRE, FORE, FBE, FOBE having temperatures of 6170~K, 6900~K, 6900~K, 7590~K respectively. Since the edge effective temperatures can depend on luminosity, these indicative effective temperatures were chosen near the center of the distribution in luminosity of the variable stars at each of the edges. Estimates of the uncertainty in these temperatures are from 150 to 200~K with the larger uncertainty for the higher temperatures. More recently \cite{Cacciari-2005} have determined effective temperatures for RR Lyrae variables in M3 from which we find very similar effective temperatures for the instability strip as those from \citeauthor{Sandage-1990}'s data, of 6150~K, 6900~K, 6700~K, and 7400~K respectively for the edges with estimates for the uncertainty in temperature of individual stars of about 100~K. One difference of note between \citeauthor{Sandage-1990} and \citeauthor{Cacciari-2005} is that the variables from \citeauthor{Cacciari-2005} form a region with a width of about 200~K where the first overtone and fundamental modes overlap while \citeauthor{Sandage-1990}'s data does not clearly indicate the presence of such a region.

Another important observation for validation of theoretical models is the variation of pulsation amplitude of RR Lyrae variables across the instability strip. For fundamental mode pulsators, cooler stars have lower radial pulsation amplitudes, and hotter stars have higher amplitudes. For first overtone mode pulsators, the amplitude of the mode forms a ``bell'' shape \citep[see Figure~\ref{fig:AV-v-temp} and also][Figures 9--11 for a theoretical depiction]{Bono-1997b}. 

Finally observed light curves can be compared with model light curves. A large number of high quality RR Lyrae light curves are available form \cite{Corwin-2001} which have been previously studied by \cite{Marconi-2009} who, using their 1D non-linear code, matched some of the available light curves. Thus the choice of comparison with M3 allows us to compare our observed light curves with many observed curves, and at the same time with 1D models. The comparison of light curves and in particular those of the cooler RR Lyrae variables near the red edge of the instability strip will help to verify our approach to modeling time dependent convection as convection plays a larger role in these stars.

\subsection{Previous Models of RR Lyrae Variables}
\label{sec:previous-models}
In our previous paper \citep[][hereafter referred to as Paper I]{Geroux-2011} we outlined the current state of modeling RR Lyrae variables, we will briefly touch on some of the more important points here. 

\cite{Gehmeyr-1992a} used a version of Stellingwerf's time-dependent convective model \citep{Stellingwerf-1982a} with an adaptive scheme and was able to produce a red edge at roughly the observed effective temperature. He notes that the effective temperature of the red edge is dependent on the parameters used for the convective model, and that the predicted temperature of the red edge could vary by a few hundred degrees Kelvin depending on the values used for the convective model parameters. He did have some difficulties, however; for example the theoretical light amplitude-rise time relation he found was different in both slope and zero point from the observed relation.

More recently \cite{Marconi-2003} used the 1D, Lagrangian, hydrodynamics code described by \cite{Bono-1994} and \cite{Bono-1997a,Bono-1997b} to compute RR Lyrae models to compare with the RR Lyrae stars observed in M3. In order to fully specify the problem \citeauthor{Marconi-2003} needed to choose a mixing-length parameter, and adopted both $l/H_p=1.5$ and $2.0$. They found that in order to match the boundaries of RR Lyrae gap in M3, they required two different mixing-length parameters, one to obtain the observed blue edge location ($l/H_p\approx 1.5$) and the other to obtain the observed red edge location ($l/H_p\approx 2.0$). In addition the observed visual amplitude as a function of $B-V$ displays nonlinear characteristics, while theoretical relations predict linear relationships. \citeauthor{Marconi-2003} also mention that a mixing-length parameter of 2.0 produces luminosities for horizontal-branch models that are brighter than is observed by $\approx0.08\pm0.05$ mag. \cite{Marconi-2009} finds that their model light curves are unable to match simultaneously the shape and amplitude of a cool RR Lyrae variable near the red edge in M3.

The distillation of multi-dimensional convective phenomenon to 1D is always accompanied by extra equations and/or parameters to approximate the effects of convective motions of material in more than one spatial dimension. \cite{Deupree-1977a} approached the interaction of convection and stellar pulsation in a fundamentally different way using a two-dimensional (2D) hydrodynamic code to directly following the convective flow patterns. While \cite{Deupree-1977b,Deupree-1977c,Deupree-1980,Deupree-1985} was able to successfully determine the observed edges of the RR Lyrae instability strip, he was unable to compute full amplitude solutions because his algorithm for moving the radial coordinate allowed the radial zoning to drift over time. Consequently at later times, the radial zoning did not cover the hydrogen ionization zone adequately and the calculations were eventually numerically unreliable. The algorithm \citeauthor{Deupree-1977a} employed for the moving radial coordinate used the horizontal average of the radial velocities at a particular radius as the grid velocity. There has been some multidimensional work on the interaction between convection and radial pulsation since then \citep{Muthsam-2010b,Mundprecht-2009,Gastine-2008a,Gastine-2011}, but to date no models have been computed capable of producing light curves which could be directly compared to observations. The current work is designed to correct this. In Paper I we presented our moving grid algorithm which allows us to use the interior mass as the radial independent variable while still using the normal angular Eulerian coordinates of a multi-dimensional spherical coordinates system.

\section{ADDITIONS FOR NON-ADIABATIC MODELS}
\label{sec:non-adiabatic-models}

\label{sec:non-ad-mods}
In Paper~I we performed only adiabatic calculations to prove the principle of forcing the radial coordinate system to move in such a way to keep the mass in a given spherical shell the same throughout the calculation.  To perform full amplitude RR Lyrae simulations for comparison with observations, we must add radiative transport, (using the diffusion approximation), a more realistic equation of state, Rosseland mean opacities, and a sub-grid scale model to our numerical code, SPHERLS. In this Section we will describe how these, as well as a few other minor changes and additions were made to SPHERLS since Paper I.

\subsection{Radiation Diffusion}
Like most 1D codes, we include heating and cooling from radiation using the radiation diffusion approximation. This approximation is used in linear and non-linear codes alike. The radiation diffusion approximation is valid when the changes in density and temperature are small over the mean free path of a photon. Near the surface however, the validity of the diffusion approximation is questionable. However, \cite{Bendt-1971} found a basic similarity between multi-frequency transport and the transport from the diffusion approximation. More recently \cite{Feuchtinger-2000} compared a calculation which used the diffusion approximation against a calculation which used time dependent radiative transfer and found only small differences in the pulsation amplitude. In addition \cite{Kollath-2000} explored the use of the radiation diffusion approximation near the surface and concluded that differences between observations and models are not a result of this inadequate treatment of the radiation transport. Because we are performing calculations that have large computational requirements, including anything more sophisticated than radiation diffusion would likely appreciably increase the computational requirements of an already highly computationally intensive simulation.

The radiation terms are converted to finite difference expressions in the manner discussed in Paper I Section 3.2, and, except for the opacity, use straight averages to compute interface quantities from zone center quantities and vice versa. To obtain the opacities at interfaces an average between adjacent, zone centered, opacities weighted by the inverse of the flux \citep{Christy-1964} as
\begin{equation}
\frac{1}{\kappa_{i+1/2}}=\frac{\frac{T_{i+1}^4}{\kappa_{i+1}}+\frac{T_{i}^4}{\kappa_{i}}}{T_{i+1}^4+T_i^4}
\end{equation}
is used, where $i$ indicates the radial zone. Another exception to the method described in Section~3.2 of Paper I is that various quantities used in the energy equation are now properly time centered. In particular $T^{n+1/2}=\frac{1}{2}\left(T^{n+1}+T^{n}\right)$ is used instead of $T^{n}$ , where $n$ denotes the time step, and the pressure, opacity, and energy, which depend on the temperature through the equation of state (described in Section~\ref{sec:equation of state}), are treated in the same way. The use of $T^{n+1}$, as it is initially unknown, is made possible by an implicit solution described in Section~\ref{sec:imp_sol}. Finally these terms also affect how the initial model is generated, as presented in the next Section.

\subsection{Changes to Starting Model Generation}
In Section~3.1 of Paper I we described the procedure for generating the starting model considering hydrostatic equilibrium for the adiabatic case. With the addition of the radiation diffusion terms to the energy equation an additional constraint can be applied to the starting model, by requiring that it also be in thermal equilibrium so that the luminosity is constant. This additional constraint removes the requirement for an energy profile from an outside source and greatly changes the procedure for generating the spherical starting model. To create a starting model including this new constraint the hydrostatic and thermal equilibrium constraints are applied to the conservations laws, resulting in
\begin{equation}
\label{eq:hydrostatic}
\frac{GM_r}{4\pi r^4}+\frac{\partial P}{\partial M_r}=0
\end{equation}
and
\begin{equation}
\label{eq:thermal-equilibrium}
\frac{-64\pi^2\sigma r^4}{3 \kappa}\frac{\partial T^4}{\partial M_r}-L=0,
\end{equation}
where $L$ is the static model luminosity. These two equations are used to define the structure of the static, fully radiative initial model by starting at the surface with boundary conditions and integrating inward in a way that is consistent with the conservation equations. The temperature in the surface zone is determined using the relation $T^4=\frac{3}{4}T_{\rm eff}^4\left(\tau +2/3\right)$ at an optical depth $\tau=0$ so that 
\begin{equation}
\label{eq:suf-temp}
T_N=2^{-1/4}T_{\rm eff}
\end{equation}
where the subscript $N$ indicates the surface zone. The mass zoning is specified by choosing an initial $\Delta M_r$. The $\Delta M_r$ for the next zone is then the initial $\Delta M_r$ increased by a chosen percentage. This percentage can be changed at various temperatures within the model to achieve a zoning tailored to capture the steep gradients in the hydrogen ionization zone while keeping the total number of radial zones reasonable. Our models typical have about 150 radial zones, about 10 of which are above the photosphere ($\tau<2/3$). We performed a few cases with about 300 radial zones and found little difference. One might be tempted to set the mass outside the model to zero, $\left(\Delta M_r\right)_{N+1}=0$, in practice the starting models have densities of about $10^{-10}-10^{-11}$ gm cm$^{-3}$ in the surface zone, while real stars have atmospheres with densities profiles that fall off with exponential tails toward zero, so that there is some additional mass outside the model at densities lower than the surface zone. To account for this $\left(\Delta M_r\right)_{N+1}=\alpha\left(\Delta M_r\right)_{N}$ with $\alpha=0.2$ similar to 1D non-linear calculations by \cite{Stellingwerf-1975} and \cite{Gehmeyr-1992a}.

The surface radius is calculated using the Stefan-Boltzmann law from the temperature and luminosity. The pressure in the surface zone, $P_N$, is calculated from Equation~(\ref{eq:hydrostatic}). Since $r$ and $M_r$ are known at the outer boundary and $\partial M_r=\frac{1}{2}\left[ \left( \Delta M_r \right)_N+\left( \Delta M_r \right)_{N+1} \right]$ is known, the only unknown is $\partial P$. $P_N$ is determined from $\partial P$ by setting the pressure at the outer interface to zero. At the outer interface of the surface zone, with our finite difference scheme, $P_{N+1/2}=\frac{1}{2}\left(P_N+P_{N+1}\right)=0$ so that $P_N=-P_{N+1}$ and $\partial P= P_N-P_{N+1}=2P_N$. We determine the density from the known temperature and pressure in the surface zone by iterating in the equation of state. The specific internal energy is obtained from the equation of state table and the opacity is obtained from another table for the known density and temperature.

Once the surface zone is fully specified, equations~\ref{eq:hydrostatic} and \ref{eq:thermal-equilibrium} are integrated into the star until the desired depth is reached, determined by reaching a chosen temperature (usually 3 million K, about $r=0.1R_{N+1/2}$).

\subsection{Equation of State and Opacities}
\label{sec:equation of state}
This work primarily uses the OPAL opacities \citep{Iglesias-1996} in combination with low temperature opacities from \cite{Alexander-1994}. For comparison purposes the King Ia table, \citep{Cox-1976b}, opacities are also used. The equation of state in all cases is from \cite{Rogers-1996}. 2D cubic spline interpolation in the $\log$ is used to create a fine rectangular grid of opacities and equation of state variables to facilitate fast linear interpolation by SPHERLS.

\subsection{Implicit Solution}
\label{sec:imp_sol}
The first nonlinear calculation of an initial value problem for a radially pulsating RR Lyrae star was performed by \cite{Christy-1964}, who solved the energy equation implicitly. \citeauthor{Christy-1964} realized that the implicit solution of the energy equation was required for the calculations to be run stably with a reasonable time step. For an explicit calculation, which involves only the nearest neighboring zones, the time step must be sufficiently small that energy change physically comes from only those nearest neighbors. With our use of the radiation diffusion approximation, this means that energy flow due to radiation can traverse only one zone in a time step. In the optically thin zones, the diffusion rate is high because the opacity is so low, producing an impractically short time step. The way around this serious restriction is to use an implicit method which couples the changes in a particular zone to the changes in all zones by solving the coupled set of finite difference energy conservation equations for all zones simultaneously.

The procedure for the implicit method is described in detail in \cite{Geroux-2013} but in summary we are using a Newton-Raphson technique to solve for the temperatures at the new timestep. We start with a guess for the temperature at the new time step, expand the finite difference expression in a Taylor series in the new temperature in every zone, keeping only the linear terms, solve the linearized equations, and repeat the process until the linear corrections are below some threshold. We have chosen the threshold so that all the relative corrections are smaller than $5\times10^{-14}$, near machine precision. Solving the linear equations is done using the Krylov subspace method, implemented in the PETSc library \citep{petsc-web-page}. This allows one to solve a large system of linear equations, quickly and accurately utilizing the message passing interface (MPI) to solve a single system using many processors at once.

\subsection{Subgrid Scale Turbulence Model}
A subgrid scale turbulence model is used to convert kinetic energy at the small scale back into thermal energy. In nature this is done by intermolecular forces or viscosity. In stars this conversion would happen on scales far smaller than those of the computational grid by many orders of magnitude. Subgrid scale turbulence models make the assumption that the large scale flows can be used to model how the subgrid scale feeds back to the large scale. In practice there are many variations on how one can do this. Since our calculations are computationally limited we choose to use a simple model based on work by \cite{Cloutman-1991} and described in \cite{Deupree-1996} and similar to \cite{Scannapieco-2008}. \citeauthor{Scannapieco-2008} used a similar model in a completely different regime to model active galactic nucleus driven turbulence in galaxy clusters, with the main difference between his model and that of \citeauthor{Cloutman-1991} being the addition of a transport equation for the eddy length scale. 

With these assumptions, the final forms of the momentum and energy conservation equations, including the radiation diffusion and the sub-grid-scale model, in spherical coordinates are
\begin{eqnarray}
\label{eq:rad-mom-cons-final}
\frac{\partial v_r}{\partial t}&+&4\pi r^2\langle\rho\rangle\left(v_r-v_{0r}\right)\frac{\partial v_r}{\partial M_r}+\frac{v_\theta}{r}\frac{\partial v_r}{\partial \theta}+\frac{v_\phi}{r\sin\theta}\frac{\partial v_r}{\partial\phi}\nonumber \\
{}&=&\frac{-4\pi r^2\langle\rho\rangle}{\rho}\frac{\partial P}{\partial M_r}+\frac{v_\theta^2}{r}+\frac{v_\phi^2}{r}-\frac{GM_r}{r^2}\nonumber \\
{}&+&\left(\frac{1}{r^2}\frac{\partial}{\partial r}\left(r^2\tau_{rr}\right)
+\frac{1}{r\sin\theta}\frac{\partial}{\partial \theta}\left(\sin\theta\tau_{\theta r}\right)\right.\nonumber \\
{}&+&\left. \frac{1}{r\sin\theta}\frac{\partial}{\partial \phi}\left(\tau_{\phi r}\right)
-\frac{\tau_{\theta\theta}+\tau_{\phi\phi}}{r}\right),
\end{eqnarray}

\begin{eqnarray}
\label{eq:theta-mom-cons-final}
\frac{\partial v_\theta}{\partial t}&+&4\pi r^2\langle\rho\rangle\left(v_r-v_{0r}\right)\frac{\partial v_\theta}{\partial M_r}+\frac{v_\theta}{r}\frac{\partial v_\theta}{\partial \theta}+\frac{v_\phi}{r\sin\theta}\frac{\partial v_\theta}{\partial\phi}\nonumber \\
{}&=&\frac{-1}{r\rho}\frac{\partial P}{\partial \theta}+\frac{v_\phi^2\cot\theta}{r}-\frac{v_r v_\theta}{r} +\left(\frac{1}{r^2}\frac{\partial}{\partial r}\left(r^2\tau_{r\theta}\right)\right.\nonumber \\
{}&+&\left.\frac{1}{r\sin\theta}\frac{\partial}{\partial \theta}\left(\sin\theta\tau_{\theta\theta}\right)
+\frac{1}{r\sin\theta}\frac{\partial}{\partial \phi}\left(\tau_{\phi\theta}\right)
+\frac{\tau_{\theta r}}{r}\right. \nonumber \\
{}&-&\left. \frac{\cot\theta\tau_{\phi\phi}}{r}\right),
\end{eqnarray}

\begin{eqnarray}
\label{eq:phi-mom-cons-final}
\frac{\partial v_\phi}{\partial t}&+&4\pi r^2\langle\rho\rangle\left(v_r-v_{0r}\right)\frac{\partial v_\phi}{\partial M_r}+\frac{v_\theta}{r}\frac{\partial v_\phi}{\partial \theta}+\frac{v_\phi}{r\sin\theta}\frac{\partial v_\phi}{\partial\phi}\nonumber \\ 
{}&=&\frac{-1}{\rho r\sin\theta}\frac{\partial P}{\partial \phi}-\frac{v_rv_\phi}{r}-\frac{v_\theta v_\phi \cot\theta}{r} +\left(\frac{1}{r^2}\frac{\partial}{\partial r}\left(r^2\tau_{r\phi}\right)\right.\nonumber \\
{}&+&\left.\frac{1}{r\sin\theta}\frac{\partial}{\partial \theta}\left(\sin\theta\tau_{\theta\phi}\right)+\frac{1}{r\sin\theta}\frac{\partial}{\partial \phi}\left(\tau_{\phi\phi}\right)+\frac{\tau_{\phi r}}{r}\right.\nonumber \\
{}&+&\left.\frac{\cot\theta\tau_{\phi\theta}}{r}\right),
\end{eqnarray}
and

\begin{eqnarray}
\label{eq:E-cons-final}
\frac{\partial E}{\partial t}&+&4\pi r^2\langle\rho\rangle\left(v_r-v_{0r}\right)\frac{\partial E}{\partial M_r}+\frac{v_\theta}{r}\frac{\partial E}{\partial\theta}+\frac{v_\phi}{r\sin\theta}\frac{\partial E}{\partial\phi}\nonumber \\
{}&+&\frac{4\pi\langle\rho\rangle P}{\rho }\frac{\partial \left(r^2v_r\right)}{\partial M_r}+\frac{P}{\rho r\sin\theta}\frac{\partial \left(v_\theta\sin\theta \right)}{\partial \theta}+\frac{P}{\rho r\sin\theta}\frac{\partial v_\phi}{\partial \phi}\nonumber \\
{}&=&\frac{4\sigma}{3\rho}\left[4\pi\langle\rho\rangle\frac{\partial}{\partial M_r}\left(\frac{4\pi\langle\rho\rangle r^4}{\kappa \rho}\frac{\partial T^4}{\partial M_r}\right)\right.\nonumber \\
{}&+&\frac{1}{r\sin\theta}\frac{\partial}{\partial \theta}\left(\frac{\sin\theta}{\kappa \rho r}\frac{\partial T^4}{\partial\theta}\right)+\left.\frac{1}{r\sin\theta}\frac{\partial}{\partial\phi}\left(\frac{1}{\kappa\rho r\sin\theta}\frac{\partial T^4}{\partial \phi}\right)\right]\nonumber \\
{}&+&4\pi\langle\rho\rangle\frac{\partial}{\partial M_r}\left(\frac{4\pi\langle\rho\rangle r^4\mu_t}{\rho Pr_t}\frac{\partial E}{\partial M_r}\right)\nonumber\\
{}&+&\frac{1}{r\sin\theta}\frac{\partial}{\partial \theta}\left(\frac{\sin\theta\mu_t}{\rho Pr_t r}\frac{\partial E}{\partial\theta}\right)\nonumber\\
&+&\frac{1}{r\sin\theta}\frac{\partial}{\partial\phi}\left(\frac{\mu_t}{\rho Pr_t r\sin\theta}\frac{\partial E}{\partial \phi}\right)+\frac{D_t\mathcal{K}^{3/2}}{L}
\end{eqnarray}
where
\begin{eqnarray}
\tau_{rr}&=&2\mu_t\left(\frac{\partial v_r}{\partial r}-\frac{1}{3}\nabla\cdot \vec{v}\right),\\
\tau_{\theta\theta}&=&2\mu_t\left(\frac{1}{r}\frac{\partial v_\theta}{\partial \theta}+\frac{v_r}{r}-\frac{1}{3}\nabla\cdot \vec{v}\right),\\
\tau_{\phi\phi}&=&2\mu_t\left(\frac{1}{r\sin\theta}\frac{\partial v_\phi}{\partial \phi}+\frac{v_r}{r}+\frac{v_\theta\cot\theta}{r}-\frac{1}{3}\nabla\cdot \vec{v}\right),\\
\tau_{r\theta}&=&\tau_{\theta r}=\mu_t\left(\frac{1}{r}\frac{\partial v_r}{\partial \theta}+\frac{\partial v_\theta}{\partial r} -\frac{v_\theta}{r}\right),\\
\tau_{r\phi}&=&\tau_{\phi r}=\mu_t\left(\frac{1}{r\sin\theta}\frac{\partial v_r}{\partial \phi}+\frac{\partial v_\phi}{\partial r}-\frac{v_\phi}{r}\right),\\
\tau_{\theta\phi}&=&\tau_{\phi\theta}=\mu_t\left(\frac{1}{r\sin\theta}\frac{\partial v_\theta}{\partial \phi}+\frac{1}{r}\frac{\partial v_\phi}{\partial \theta}-\frac{v_\phi\cot\theta}{r}\right)
\end{eqnarray}
and
\begin{equation}
\label{eq:smag-edd}
\mu_t=\frac{C^2l^2\rho}{\sqrt{2}}\left(\nabla\vec{v}:\left[\nabla\vec{v}+\left(\nabla\vec{v}\right)^{T}\right]\right)^{1/2}.
\end{equation}
In the above equations $\langle\rho\rangle$ denotes the horizontal average density and $\mathcal{K}$ is the subgrid scale turbulent kinetic energy density. Equation~(\ref{eq:smag-edd}), from \cite{Smagorinsky-1963}, is the simplest method to calculate an eddy viscosity coefficient where $l$ is the length scale of a grid zone, and $C$ is a constant of approximately 0.17 based on comparison between calculations and experiments by \cite{Deardorff-1971}. The expression $\mu_t=A_t\rho L\mathcal{K}^{1/2}$ from \cite{Cloutman-1991}, where $L=3.75\times l$ and $A_t$ is a constant of 0.117, is used to solve for $\mathcal{K}$ from $\mu_t$. The values of the constants $D_t=1.4$, $Pr_t=0.7$ (the turbulent Prandtl number) are from \cite{Cloutman-1991}. Equations~(\ref{eq:rad-mom-cons-final})~--~(\ref{eq:E-cons-final}) with the mass conservation equation given in Paper I represent the final form of the conservation equations used.

\section{1D MODELS}
\label{sec:1D-models}
As a precursor to the multi-dimensional calculations we have calculated a number of 1D (radiation only) models. These models serve two major purposes. They were helpful in examining the parameter space or model properties, and they provide a fiducial point to compare with the 2D and three-dimensional (3D) convective models. A basic set of 1D calculations used for general information is summarized in Table~\ref{table:1D-models}. Further 1D models were computed to compare with specific 2D and 3D calculations as needed and are presented in the next Section. Each calculation set in Table~\ref{table:1D-models} contains a number of models computed at temperatures of 6000, 6100, 6200, 6300, 6500, 6700 and 6900 K.  All calculations have the same composition of $X=0.7$ and $Z=0.001$, which was chosen to match the composition of the King Ia opacity table \citep{Cox-1976b}. All calculations use the OPAL equation of state from \cite{Rogers-1996}. The calculations were computed at two different masses (0.575 and 0.7 $M_\odot$), with two different opacity tables (King Ia opacity table and the OPAL opacity table; \citep{Iglesias-1996}), and with two different initial surface velocities for the radial pulsation velocity profiles. The relative radial velocity profiles were taken form linear, adiabatic calculations of the radial eigenfunctions made with the LNA code \citep{Castor-1971}.

\begin{deluxetable}{ccc}
\tablecaption{One-dimensional calculation set parameters. Each set has models calculated at temperatures of 6000, 6100, 6200, 6300, 6500, 6700 and 6900 K and a composition of X=0.7, and Z=0.001 \label{table:1D-models}}
\tablehead{\colhead{$\left(u_0\right)_{\rm surf. init.}$} & \colhead{Opacity table} & \colhead{Mass} \\ 
\colhead{(km s$^{-1}$)} & \colhead{} & \colhead{(M$_\odot$)} }
\startdata
2 & KING Ia & 0.575\\
2 & OPAL & 0.575\\
2 & KING Ia & 0.7\\
2 & OPAL & 0.7\\
10 & OPAL & 0.7\\
\enddata
\end{deluxetable}

\begin{figure}

\center
\plotone{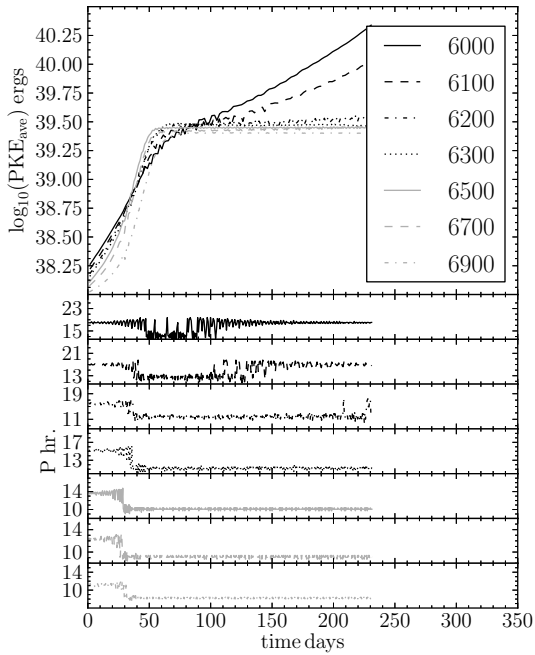}
\caption{The upper panel shows the log of peak kinetic energy while the lower panel shows the periods in hours. This figure is for the calculation set using OPAL opacities, a mass of M=0.7M$_\odot$, and with the radial pulsation initiated at 2~km~s$^{-1}$.}
\label{fig:OPAL-M7}
\end{figure}

\begin{figure}

\center
\plotone{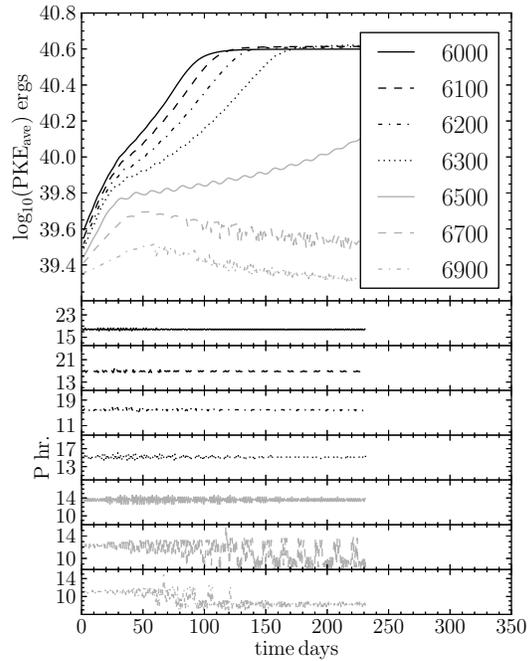}
\caption{Similar to figure~\ref{fig:OPAL-M7} except the radial pulsation was initiated at 10~km~s$^{-1}$ instead of 2~km~s$^{-1}$.}
\label{fig:OPAL-M7-highv}
\end{figure}

The peak kinetic energy of the radial pulsation is generally used for measuring the growth or decay of a mode. The cessation of peak kinetic energy growth indicates when the radial pulsation reaches full amplitude. Figures~\ref{fig:OPAL-M7} and \ref{fig:OPAL-M7-highv} show the growth of the peak kinetic energy for models starting at a low initial pulsation amplitude and at relatively large initial pulsation amplitude, respectively. The top panels of these figures show the growth of the peak kinetic energy averaged over three fundamental periods. The average is meant to remove variations due to contamination from the first overtone mode as this time interval is approximately four first overtone periods \citep{Christy-1964}. Though these calculations are initiated in the fundamental mode, the initial linear mode is not strictly correct at higher non-linear amplitudes and contamination from the first overtone and probably higher overtones occurs. High frequency contamination usually disappears quickly as evident by the loss of the high frequency noise in the light curves.

The bottom panels of figures~\ref{fig:OPAL-M7} and \ref{fig:OPAL-M7-highv} show the period as a function of time, as determined from the time between every other peak in kinetic energy (there are two peaks in the kinetic energy per period, one for expansion and one for contraction). Interestingly we find that the models switch radial pulsation modes. From the periods it is clear when the mode changes from the fundamental to the first overtone, as there is a large drop in the period to about 0.72 times the fundamental period. The mode can also be discerned from the magnitude of the peak kinetic energy when it levels off, although this is not as direct an indicator as the period. It should be noted that many models, particularly those at low temperature, that change from the fundamental to the first overtone also change back to the fundamental at later time. The time at which this second transition begins is later for higher temperature models.

From examining these figures and similar figures for the other simulation sets listed in Table~\ref{table:1D-models}, three main observations are noted. First, for 0.575~M$_\odot$ models growing from a low amplitude radial pulsation, the King Ia and OPAL tables have very different temperatures at which they switch from the fundamental to the first overtone mode. The radial pulsation calculations using the OPAL tables all grew into the first overtone in the range of  temperatures explored, while the calculations using the King Ia table grew into either the fundamental or first overtone with the switch between the two modes occurring between 6500 (fundamental) and 6700 (first overtone). Second, at higher mass, for models started at low amplitude, the switch between the first overtone and fundamental modes occurs at about the same temperature between 6200 (fundamental) and 6300 (first overtone) for both the King and OPAL opacity tables. Finally, for models starting at a higher amplitude, which encourages the mode to remain the fundamental mode it started in more than the low amplitude calculations, the temperature at which the mode switches is increased (compare figures~\ref{fig:OPAL-M7} and \ref{fig:OPAL-M7-highv}). For the OPAL tables with a mass of 0.7~M$_\odot$ starting with an initial surface radial velocity of 2~km~s$^{-1}$, the switch between modes occurs between 6100 (fundamental) and 6200 (first overtone). For the calculations initiated at 10~km~s$^{-1}$ the switch occurs between 6500 and 6700~K. If calculations are initiated at even higher radial velocities of 20~km~s$^{-1}$ the 6700~K model remains in the fundamental mode, while the 6900~K model decays and would likely end up in the first overtone if given enough time. 

The causes for these various mode switches depend at least on the model effective temperature, model mass, the model initial velocity amplitude, and the details of the opacity table. There is no reason to believe, while interesting, that these are the only properties which influence this behaviour. Further exploration of this phenomenon is beyond the scope of this work. However, these studies do allow us to determine the radial pulsation amplitude to begin with for each temperature and opacity to arrive at full amplitude in the least amount of computational time, something of interest in the 2D and 3D simulations.

\section{Interaction of Convection and Pulsation}
Here we extend the work to 2D models. These models include convection using two dimensional versions of the conservation equations together with the constitutive and radial grid flow equations (see Paper I for details of the radial grid flow equation). This formulation of convection allows the convective flow to arise naturally from the conservation equations, albeit only in two spatial dimensions. The convective motions grow in regions unstable to convection from small deviations from the spherical model resulting from machine round off errors. In regions stable to convection and which are sufficiently far from convectively unstable regions, the gas motions, other than radial pulsation, remain small, near machine precision. The convective velocities discussed in this and the proceeding Sections are measured from the radial and $\theta$ velocities with the radial grid velocity subtracted from the radial velocity. The intent of defining the convective velocity in this way is to obtain a measure of the convective motion without including the radial pulsation motion; however, this definition of the convective motion may not strictly measure actual convective velocities because none of the motion is strictly Lagrangian.

Our computational domain covers a pie slice of 6$^o$ and extends down to more than 90\% of the star by radius. The angular size of this region was chosen from an examination of 3D simulations (to be presented in an upcoming paper) of various angular extents in which we found that 6$^o$ is large enough for more than one 3D convective cell to form, ensuring that the convective cells may grow to their preferred size and are not limited by the extent of the simulation. In these simulations we use 20 angular zones and approximately 150 radial zones. The choice of 20 angular zones is for later comparison with 3D calculations which will have 20 angular zones in both angular directions. The number of angular zones in the 3D calculations was a compromise between completing the calculations in a reasonable amount of time and resolution. The radial zoning was chosen so that there was at least one zone in the hydrogen ionization region at all times which in turn requires quite fine zoning. As a result the cell aspect ratio at the surface is about 65. This coarse angular zoning will not be adequate to model the fine detail of the convection but appears to be adequate for modeling the net effects of convection on the radial pulsation.  The radial extent was chosen so that the details of the nuclear reactions could be ignored while still producing the correct observational radial pulsation periods.

We have computed a number of 2D models at effective temperatures ranging from 6200 to 6900~K in steps of 100~K. These models all have a mass of 0.7 M$_\odot$, a static model luminosity of 50~L$_\odot$, and use the OPAL opacities and equation of state. The composition of all 2D models was chosen to be X=0.7595, Z=0.0005. The models had radial pulsation initiated in the fundamental mode from the linear eigenfunctions scaled to have a surface radial velocity of either 10~km~s$^{-1}$ or 20~km~s$^{-1}$. The mass, luminosity, and composition were chosen to be representative of RR Lyrae variables in M3 as follows. The mass and luminosity were selected by taking an average of RR Lyrae masses and luminosities in M3 as determined by \cite{Cacciari-2005}. The composition's helium mass fraction, Y=0.24, is the same value used by \cite{Marconi-2007} for comparison of 1D models to M3's observed RR Lyrae light curves. The metal mass fraction selected was slightly larger than that assumed by \cite{Bono-1997b} of Z=0.0004 for M3 while lower than the value used by \cite{Marconi-2007} of 0.001. The goal is to compare the the full amplitude light curves obtained with this multi-dimensional approach to convection with both the observed light curves and the light curves previously computed using the mixing length theory of convection. For such a comparison modest composition variations such as these play only a minor role. Ideally one would like to compute a number of models at various compositions, masses, and luminosities for comparison with observations. However, the computational requirements to cover this large parameter space with even modest resolution (e.g. all possible combinations of two values of the four parameters of mass, luminosity, Y, and Z results in sixteen effective temperature sets) with this approach to convection is prohibitive, and beyond the scope of the present work. 

Early work by \cite{Tuggle-1973} theorized that even the presence of time independent convection would quench pulsation, although their work did not obtain negative growth rates even at very low effective temperatures. While time independent convection greatly reduced the growth rate near the red edge, it did not become negative (which is required for the decay of a mode). Later work by \cite{Deupree-1977a} showed it was not that convection carried a large portion of the flux all the time, negating the kappa mechanism, but instead it is the time dependence of convection that is important for damping pulsation. This time dependence of convection on pulsation phase does more than simply remove the driving mechanism from the ionization zone, it turns this zone into a damping region if convection carries a sufficient amount of the total flux at the right pulsation phases.

The present work also explores the time dependence by computing the convective flux, the amount of energy carried radially by convection per unit area. The approach to calculating this quantity is to assume that the amount of energy carried by convection is related to the difference between the temperature of a given cell and the horizontal average of the temperature, $\Delta T$, multiplied by the specific heat at constant pressure, $c_P$, and the mass flux across the outer radial interface of the shell. This results in
\begin{equation}
\label{eq:con-lum}
F_{{\rm conv.}}=c_P\rho \left(v_r-v_{r0}\right)\Delta T.
\end{equation}
Equation~(\ref{eq:con-lum}) may not be entirely correct as it assumes that the energy deposited or removed from the surroundings by the rising or falling material is a result of the material reaching thermal equilibrium with the surroundings. This may not be true if the material rises or falls again before it equilibrates to the horizontal average temperature. It is also assumed that the horizontal pressure variation is negligible, which, while not exact, is not a bad approximation. While Equation~(\ref{eq:con-lum}) may not be exact it should  provide a good relative measure of the energy carried by convection for comparisons among the various models. It should be noted that this equation is used in analysis only and is not part of the numerical simulation.

\begin{figure}
\center
\plotone{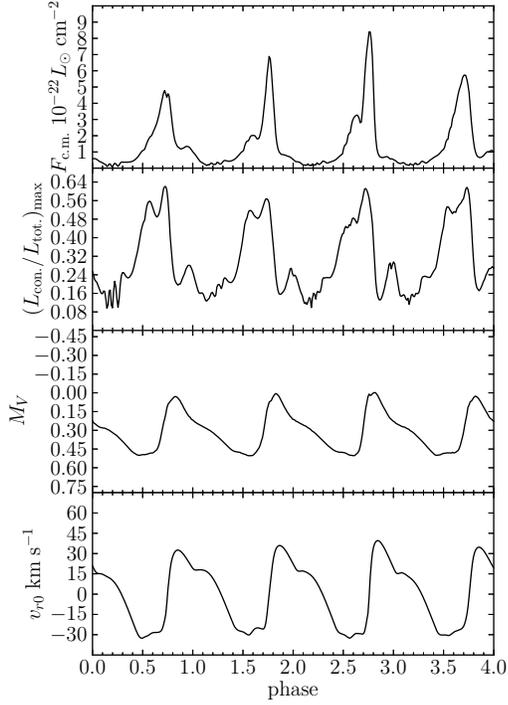}
\caption{Convective time dependence of 6300~K effective temperature model. From top to bottom: the maximum of the convective flux through out the model, the maximum over all radial shells of the ratio of the convective luminosity to the total luminosity, the absolute visual magnitude, and the surface radial grid velocity.}
\label{fig:con-time-dep-6300}
\end{figure}

\begin{figure}
\center
\plotone{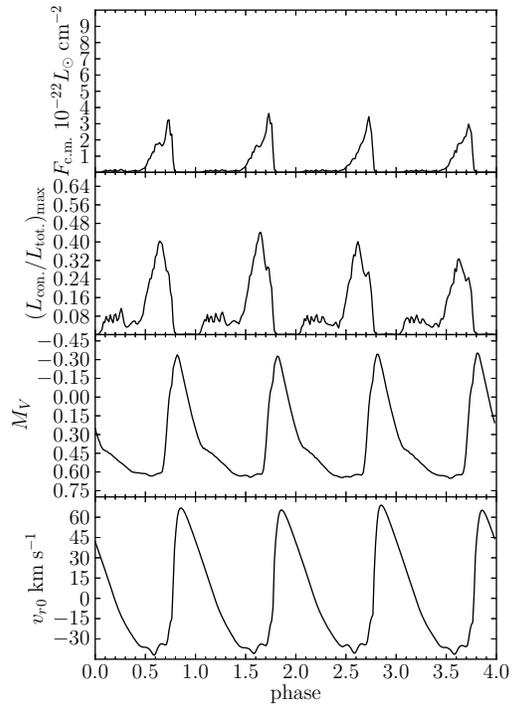}
\caption{Similar to figure~\ref{fig:con-time-dep-6300} except for the 6700~K effective temperature model.}
\label{fig:con-time-dep-6700}
\end{figure}

Figures~\ref{fig:con-time-dep-6300} and \ref{fig:con-time-dep-6700} show how the maximum convective flux throughout the model varies with time for the 6300, and 6700~K models respectively. In the second pane from the top is plotted the maximum value over all radial shells of the ratio of convective luminosity to the total luminosity. Also shown in the bottom two panels are the light and velocity curves. The determination of the bolometric corrections required to make the light curve will be discussed in the next Section. As the model contracts to minimum size, the convective flux grows. While the model begins to expand the maximum convective flux drops steeply. This is the same time dependence that \citeauthor{Deupree-1977a} found to quench pulsation at the red edge of the RR Lyrae instability strip.  It is also clear when comparing the two figures, that the maximum convective flux is greatly reduced in the hotter model so that the role of convection becomes less important as the model's effective temperature increases. It is noted from the second pane from the top of figures~\ref{fig:con-time-dep-6300}~and~\ref{fig:con-time-dep-6700} that the largest fraction of the energy transported by convection is about 64\% and 40\% for the 6300~K and 6700~K models respectively. This is significantly less than the nearly 99\% that one would infer from the standard mixing length theory. This division of the energy flow into convective and radiative components is quite different from that predicted by the local mixing length theory because the horizontally averaged opacity can be far from the opacity of the horizontally averaged temperature \citep{Deupree-1980a}, as assumed in the local mixing length theory.

\begin{figure}
\center
\plotone{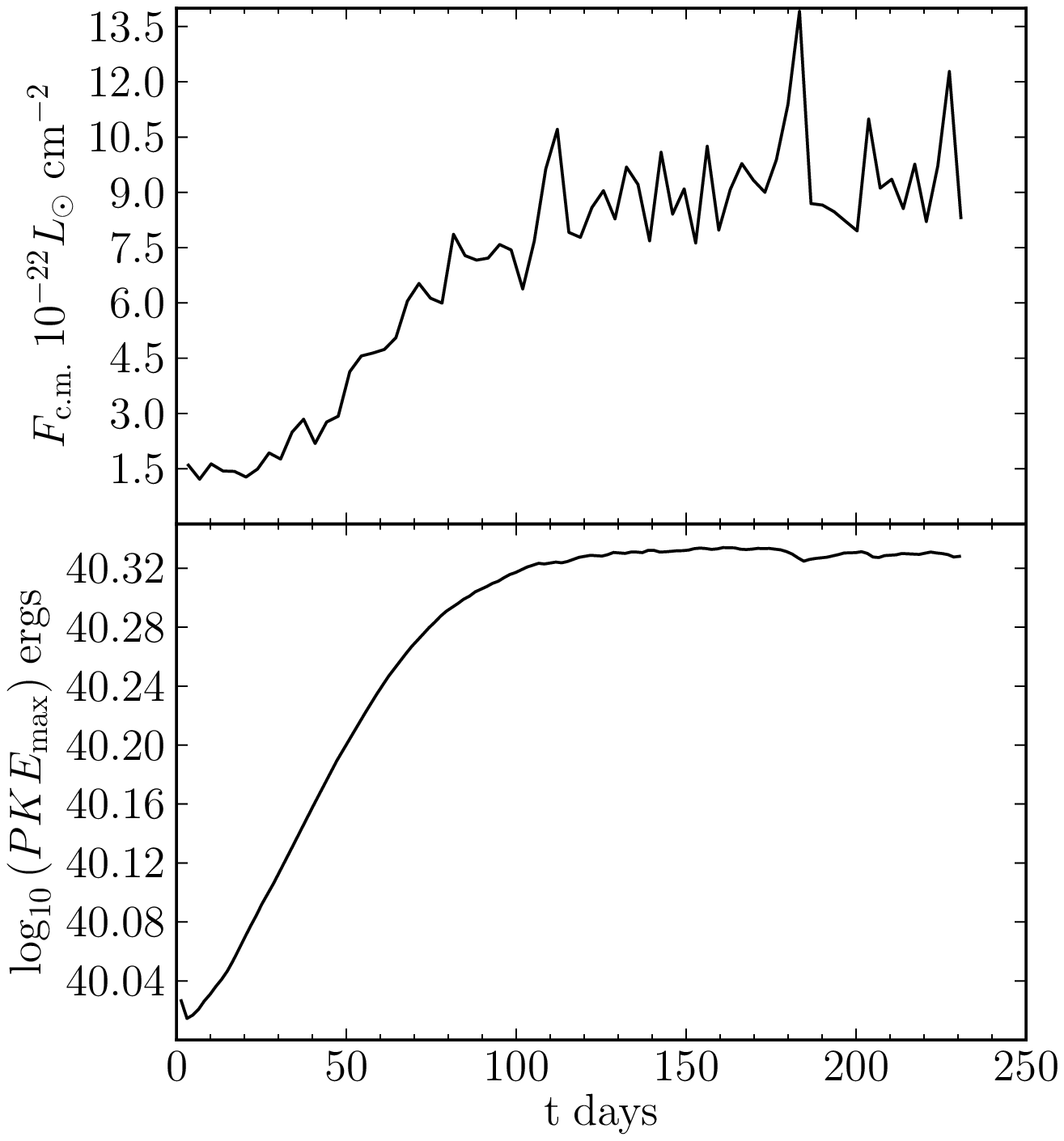}
\caption{The top panel shows the six period average of the peak convective flux for 6500~K model. The bottom panel shows the growth of the $\log$ of the peak kinetic energy averaged over three periods, for the same model.}
\label{fig:con-l-amp-dep}
\end{figure}

Interestingly, we find the strength of convection increases with increasing pulsation amplitude. We show this in figure~\ref{fig:con-l-amp-dep} for the 6500~K model. The flux in figure~\ref{fig:con-l-amp-dep} is computed in the following manner: for each time we compute the maximum convective flux in the mesh, then we find the highest value of this maximum convective flux over a single pulsation period, and finally we compute one point in figure~\ref{fig:con-l-amp-dep} by taking the average of the highest values for six consecutive periods. The average in the convective flux is mainly to remove contamination from the first overtone mode, as done in the calculation of the average peak kinetic energies. From figure~\ref{fig:con-l-amp-dep} it is obvious that as the pulsation amplitude increases, so too does the strength of convection. This is possibly a result of the hydrogen ionization zone sweeping through more of the star as the pulsation amplitude increases. Convection can then penetrate deeper and include more mass and thus convection is able to carry more energy. However, the interaction of convection and pulsation is complex and this relationship may be more complex than such a simple explanation. The maximum convective flux displayed in figure~\ref{fig:con-l-amp-dep} is the maximum throughout the model, occurring in only one radial zone, which is maintained for only a small part of the pulsation cycle (see figure~\ref{fig:con-time-dep-6300} or \ref{fig:con-time-dep-6700}) and remains very low through the rest of the pulsation cycle.

\begin{deluxetable}{cccc}
\tablecaption{Peak kinetic energy growth rates in \% increase per period for 1D fully radiative models, $\eta_{\rm 1D}$, and 2D convective models, $\eta_{\rm 2D}$, and the their difference. \label{table:growth-rates}}
\tablehead{\colhead{$T_{\rm eff}$} & \colhead{$\eta_{\rm 1D}$} & \colhead{$\eta_{\rm 2D}$} &\colhead{$(\eta_{\rm 1D}-\eta_{\rm 2D})$}\\ 
\colhead{(K)} & \colhead{} & \colhead{} & \colhead{} }
\startdata
6200 & 2.43 & 1.27 & 1.16\\
6300 & 2.23 & 1.15 & 1.08\\
6400 & 1.98 & 0.98 & 1.00\\
6500 & 1.67 & 0.74 & 0.93\\
6600 & 0.79 & 0.30 & 0.49\\
6700 & 0.93 & 0.32 & 0.61\\
6800 & 0.24 & 0.04 & 0.20\\
\enddata
\end{deluxetable}

Growth rates of the peak kinetic energy of pulsation were calculated for all 2D models and their 1D model counterparts. These rates are compared in Table~\ref{table:growth-rates}. Models used to calculate growth rates were all initiated at the same surface radial velocity of 10~km~s$^{-1}$. The 1D models given in this table were computed to match the 2D models in all respects except that the 1D models are fully radiative while the 2D models allow hydrodynamic convection. As the effective temperature becomes cooler the difference in growth rates between the fully radiative 1D models and the 2D convective models becomes larger.

We had hoped to clearly show that convection quenches pulsation at the red edge of the instability strip by noting a decrease in the peak kinetic energy (negative growth rate) of models cooler than some temperature. However, as the models became cooler convection became stronger, and eventually the convectively unstable region penetrated well below the hydrogen and helium ionization regions. Once this happened there was a net change in the global model structure and hence in the potential energy of the model. This change in potential energy would require a large amount of time (at least relative to the times covered by these simulations) to return the model to some sort of equilibrium underlying the pulsation. This adjustment significantly affects the use of the peak kinetic energy as a measurement of the growth or decay of a pulsation mode. The 6200~K model did not reach full amplitude before the convectively unstable region penetrated below the ionization zones. The 6300~K model appeared to be at nearly full amplitude just before the convection penetrated below the ionization zones, as the growth of the peak kinetic energy of the pulsation mode had slowed considerably. Models with effective temperatures 6400~K and hotter reached full amplitude modes without this deeper penetration.

\section{Full amplitude Solutions}
\label{sec:full-amp-sol}
We have computed a number of 2D models to full amplitude for the fundamental mode. These are listed in Table~\ref{table:2D-model-parameters}. The initial velocities were determined from the 1D calculation for the same static model as indicated in the previous Section.

\begin{deluxetable}{ccccccc}
\tablecaption{Properties of 2D models.\label{table:2D-model-parameters}}

\tablehead{\colhead{Model} & \colhead{$T_{\rm teff}$} & \colhead{$L$} & \colhead{$M$} & \colhead{$\Pi$} & \colhead{$A_V$} & \colhead{Mode}\\ \colhead{} & \colhead{(K)} & \colhead{$(L_\odot)$} & \colhead{$(M_\odot)$} & \colhead{(hr.)} & \colhead{} & \colhead{} }
\startdata
T6300 & 6300 & 50 & 0.7 & 15.09 & 0.56 & RRab\\
T6400 & 6400 & 50 & 0.7 & 14.32 & 0.64 & RRab\\
T6500 & 6500 & 50 & 0.7 & 13.57 & 0.75 & RRab\\
T6600 & 6600 & 50 & 0.7 & 12.89 & 0.83 & RRab\\
T6700 & 6700 & 50 & 0.7 & 12.24 & 1.01 & RRab\\
T6800 & 6800 & 50 & 0.7 & 11.63 & 0.86\tablenotemark{a} & RRab\\
T6900 & 6900 & 50 & 0.7 &  8.26 & 0.51 & RRc\\
\enddata

\tablenotetext{a}{This is a lower bound for $A_V$ as the 6800~K model has not yet reached full amplitude.}

\end{deluxetable}

Models hotter than 6400~K were initiated with 20~km~s$^{-1}$ surface pulsation velocity while models at temperatures of 6400~K and cooler were initiated at the 10~km~s$^{-1}$ surface velocity. The higher initial radial pulsation velocity was chosen because it shortened the time required to reach full amplitude, and was required by hotter models to remain pulsating in the fundamental mode. If models closer to the fundamental blue edge (6600~K and hotter) are initiated at the lower velocity, then their pulsation amplitude decays and eventually the pulsation mode switches into the first overtone pulsation mode. We note that the mode may switch back to the fundamental given sufficiently long computational time for these hotter models. This was illustrated in the 1D models in Section~\ref{sec:1D-models} and is also seen in the 2D simulations.

\begin{figure}
\center
\plotone{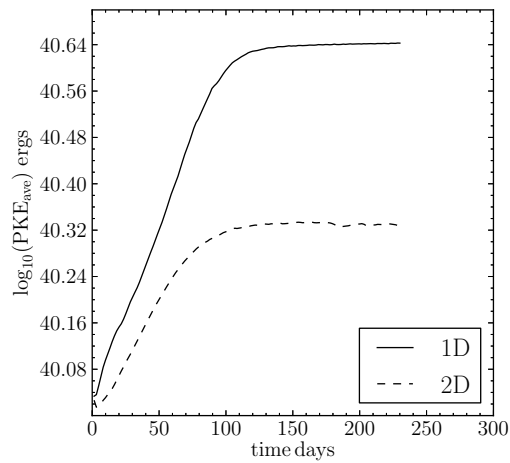}
\caption{A comparison of 1D fully radiative and 2D convective model peak kinetic energies for 6500~K models.}
\label{fig:PKE-1D-2D_comp}
\end{figure}

Even when initiated with a surface pulsation velocity of 20~km~s$^{-1}$ the pulsation amplitude of the 6900~K model decayed, and presumably it would eventually switch to the first overtone mode. At 10~km~s$^{-1}$ the pulsation amplitude also decayed and the pulsation mode switched to the first overtone mode and remained at full amplitude. The 6900~K model referenced in Table~\ref{table:2D-model-parameters} was initialized with the linear fundamental eigenfunction mode scaled to a surface velocity of 5km~s$^{-1}$. The mode slowly switched from the fundamental mode to the first overtone as the pulsation amplitude grew until it reached full amplitude in the first overtone mode.

Models cooler than 6500~K were initiated at the lower, $10$~km~s$^{-1}$ velocity to ensure that their pulsation amplitude would grow to the full amplitude rather than decay from higher amplitude to their equilibrium pulsation amplitude or full amplitude. If started at a pulsation amplitude higher than the equilibrium pulsation amplitude, the assumption of a linear eigenfunction for the mode is less valid, and contamination from other modes is more of a concern. An example of the growth of the peak kinetic energy with time is shown in figure~\ref{fig:PKE-1D-2D_comp} for a 2D convective model along with a 1D fully radiative model with otherwise similar model parameters, both with an effective temperature of 6500~K. It is clear that convection limits the amplitude of pulsation as compared to the fully radiative models.

The main reason for computing full amplitude radial pulsations is to compare computed model results with observations. To this end we have computed visual light curves from model absolute bolometric light curves using a table of bolometric corrections as functions of $\log_{10}g$ and $T_{\rm eff}$ from \cite{Bessell-1998} model atmospheres. Our model's absolute bolometric magnitudes are converted to the absolute visual magnitude at individual points in the light curve by linearly interpolating in the table of bolometric corrections to the effective temperature and gravity of our model at that point in the light curve. Spot checks in the models show that horizontal variations in the effective temperature tend to be less than 30~K. This effectively assumes that the bolometric correction in the pulsating model is the same as in a static model at the same effective temperature and gravity. Applying the bolometric correction acts to increase the amplitude of the visual light curve over the bolometric light curve, as the bolometric correction is largest at maximum light, and smallest at minimum light. This was also found by \cite{Kovacs-1998} (see their figure~3).  We have not included the acceleration from the surface velocity, as it occasionally leads to unrealistic values of the effective gravity. We have checked the difference between including and not including this acceleration in cases where it does not result in unrealistic effective gravities, and omitting it does not substantially effect the results. \citeauthor{Kovacs-1998} have shown that the error introduced by neglecting even the variation of the effective gravity with pulsation phase is negligible as the effective temperature plays the largest role in determining the variation of the bolometric correction over a pulsation cycle.

\begin{deluxetable}{cccccccc}
\tablecaption{Properties of choice M3 variables as given in \cite{Cacciari-2005} and the models for which they are the best match. \label{table:M3-variables}}
\tablehead{\colhead{Star} & \colhead{$T_{\rm teff}$} & \colhead{$L$}& \colhead{$M$}& \colhead{$\Pi$}& \colhead{$A_V$}& \colhead{Mode}& \colhead{Model} \\ 
\colhead{} & \colhead{(K)} & \colhead{$(L_\odot)$}& \colhead{$(M_\odot)$}& \colhead{(hr.)}& \colhead{}& \colhead{}& \colhead{Matched}}
\startdata
v120 & 6300 & 49.9 & 0.68 & 15.36 & 0.44 & RRab& T6300\\
v19  & 6266 & 49.3 & 0.70 & 15.17 & 0.45 & RRab& T6300\\
v48  & 6283 & 58.7 & 0.67 & 15.07 & 0.61 & RRab& T6400\\
v93  & 6446 & 50.2 & 0.67 & 14.46 & 0.73 & RRab& T6500\\
v10  & 6469 & 51.4 & 0.73 & 13.67 & 0.88 & RRab& T6600\\
v92  & 6712 & 49.9 & 0.71 & 12.07 & 1.14 & RRab& T6700\\
v125 & 6829 & 49.4 & 0.71 &  8.40 & 0.41 & RRc& T6900\\
\enddata
\end{deluxetable}

\begin{figure}
\center
\plotone{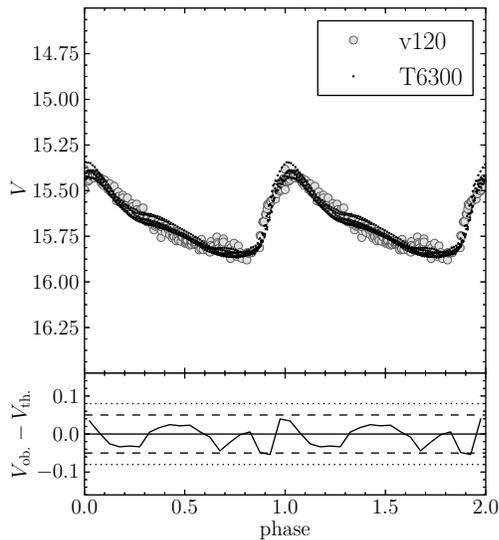}
\caption{6300~K effective temperature model's light curve compared to variable star v120's light curve using a distance modulus of 15.34. The vertical scale was chosen to match the scale of the light curve figures of \cite{Marconi-2007}.}
\label{fig:LC-T6300_v120}
\end{figure}

\begin{figure}
\center
\plotone{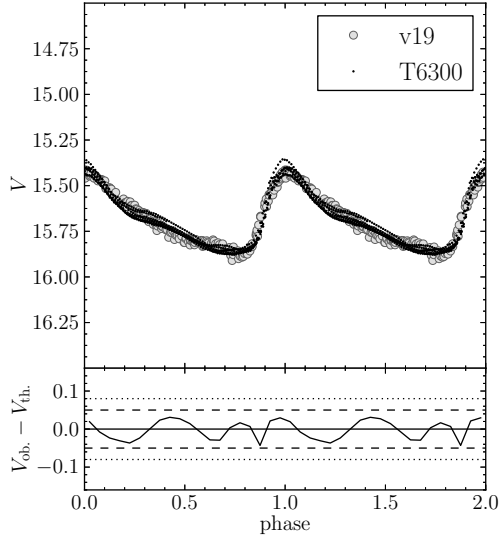}
\caption{6300~K effective temperature model's light curve compared to variable star v19's light curve using a distance modulus of 15.35.}
\label{fig:LC-T6300_v19}
\end{figure}

\begin{figure}
\center
\plotone{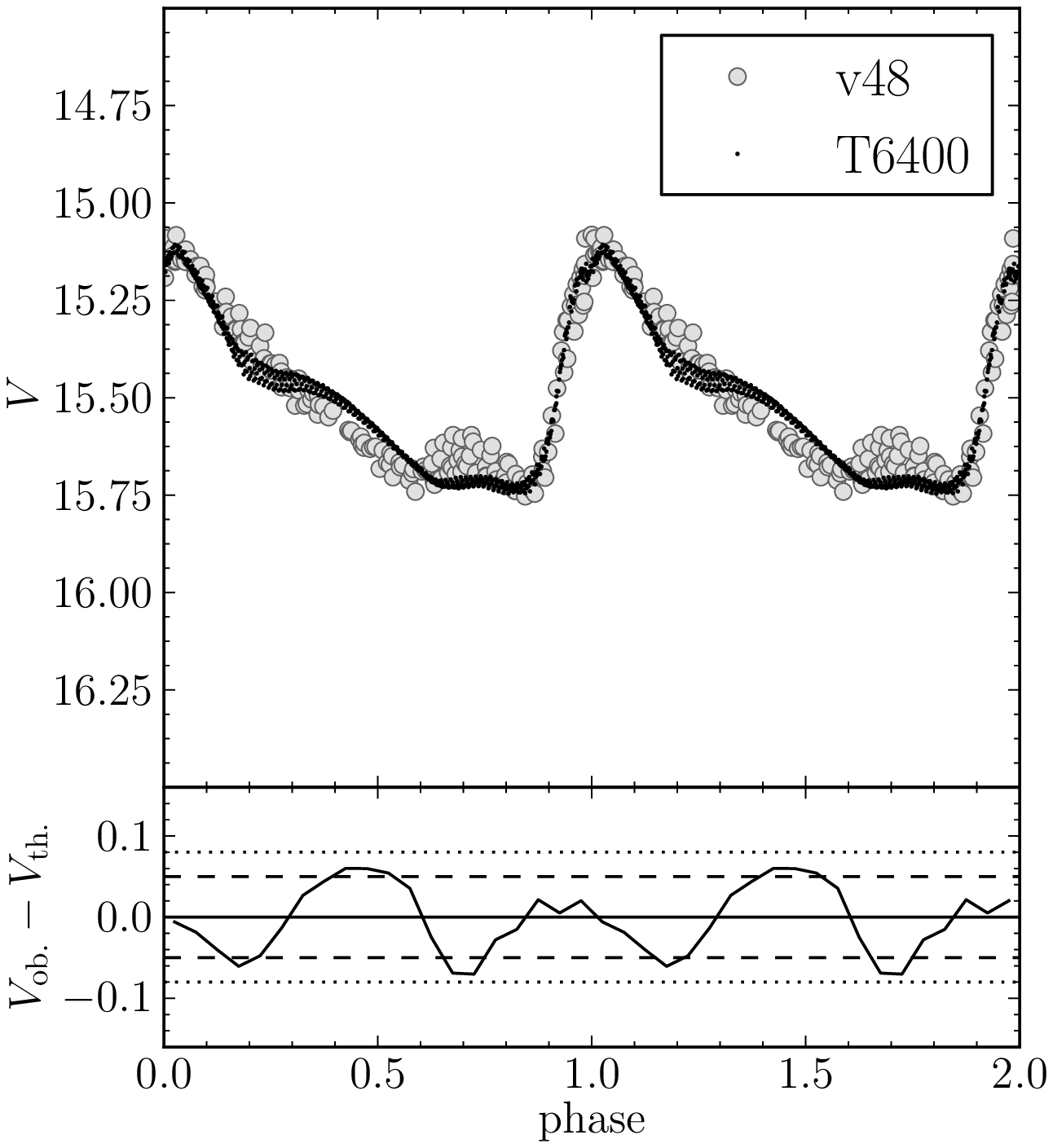}
\caption{6400~K effective temperature model's light curve compared to variable star v48's light curve using a distance modulus of 15.16.}
\label{fig:LC-T6400_v48}
\end{figure}

\begin{figure}
\center
\plotone{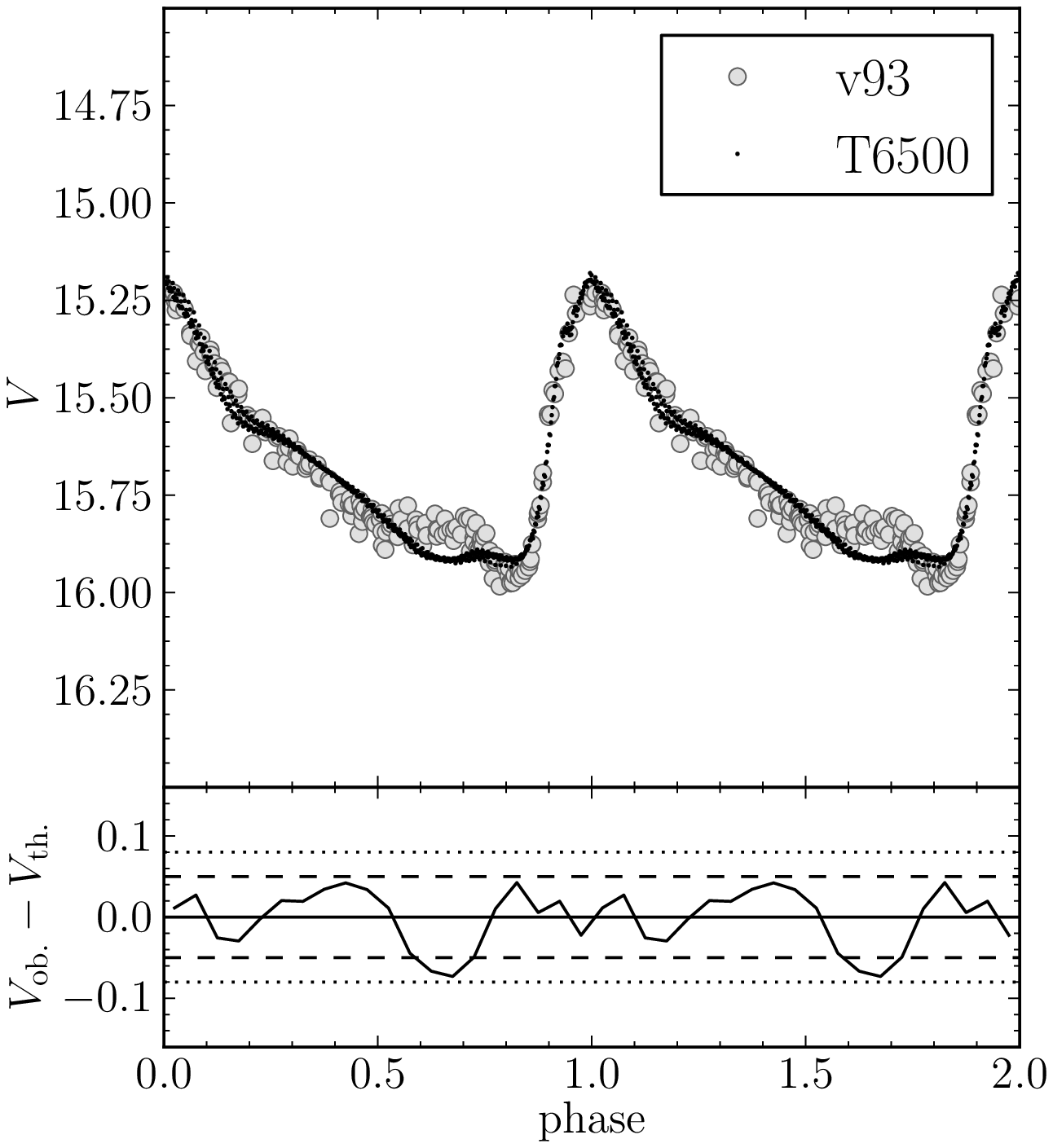}
\caption{6500~K effective temperature model's light curve compared to variable star v93's light curve using a distance modulus of 15.32.}
\label{fig:LC-T6500_v93}
\end{figure}

\begin{figure}
\center
\plotone{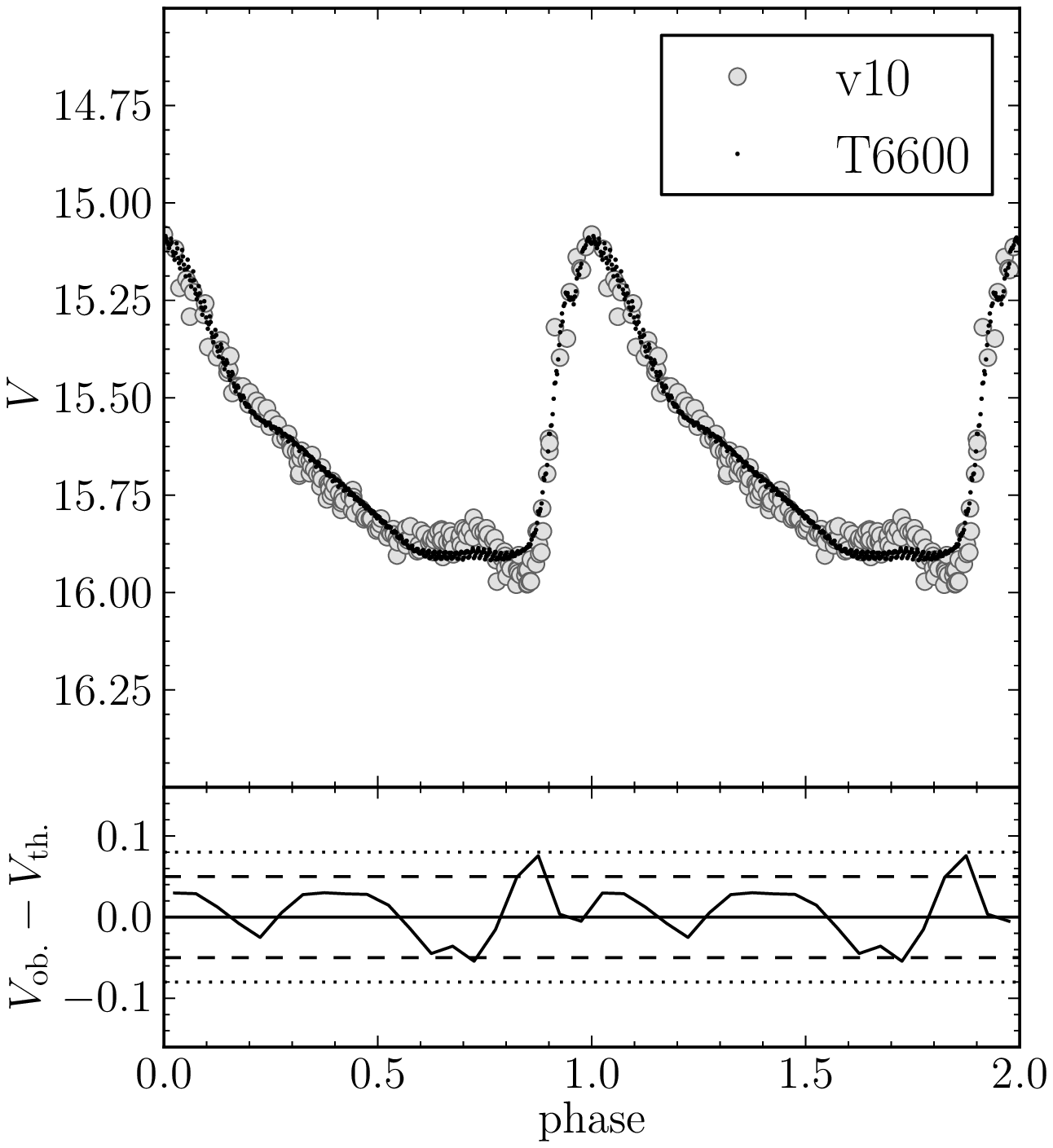}
\caption{6600~K effective temperature model's light curve compared to variable star v10's light curve using a distance modulus of 15.27.}
\label{fig:LC-T6600_v10}
\end{figure}

\begin{figure}
\center
\plotone{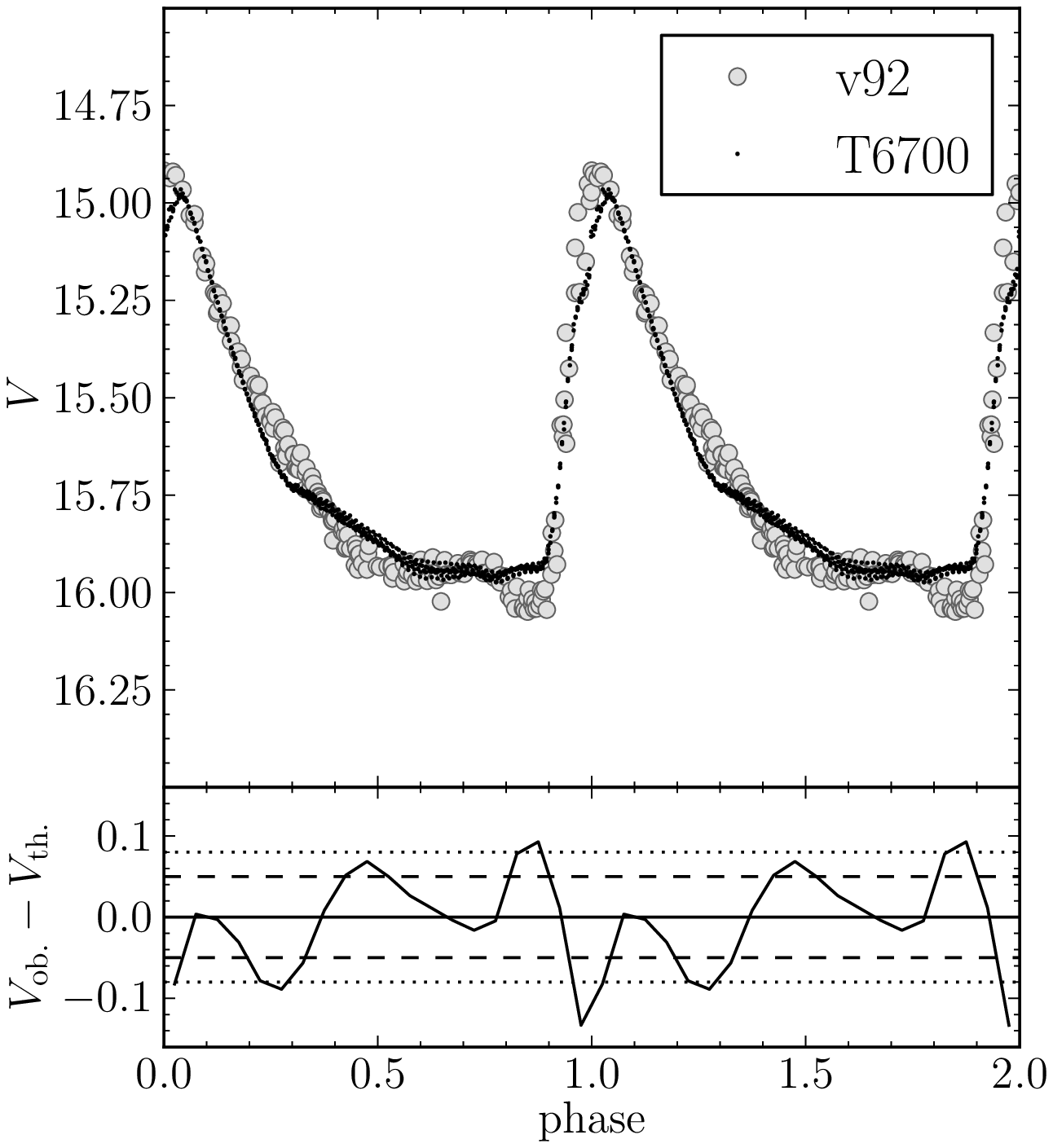}
\caption{6700~K effective temperature model's light curve compared to variable star v92's light curve using a distance modulus of 15.32.}
\label{fig:LC-T6700_v92}
\end{figure}

\begin{figure}
\center
\plotone{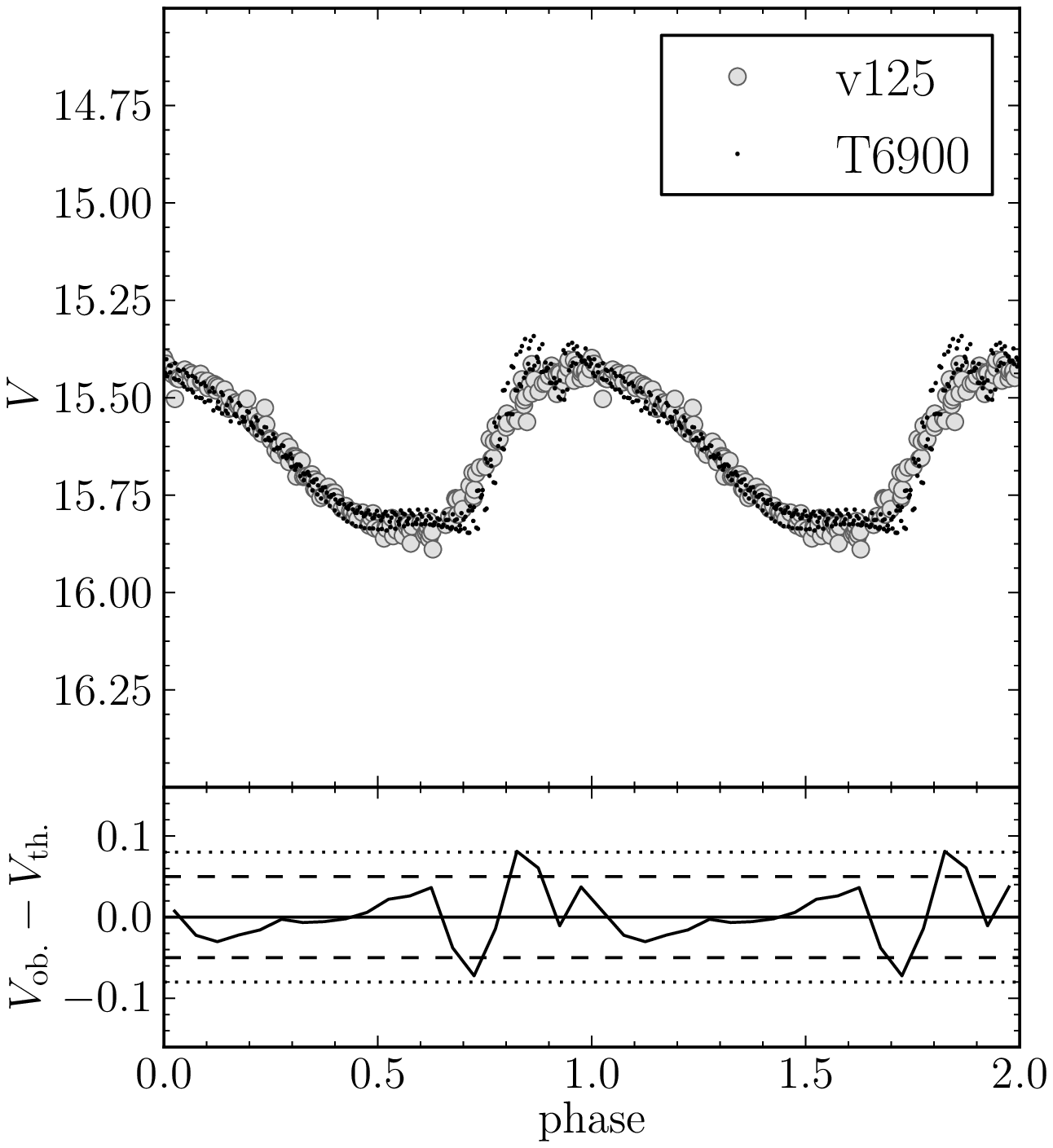}
\caption{6900~K effective temperature model's light curve compared to variable star v125's light curve using a distance modulus of 15.32.}
\label{fig:LC-T6900_v125}
\end{figure}

We have computed a number of synthetic visual light curves in the way described above. The 2D models provide absolute bolometric magnitudes at regularly spaced time intervals of 600 seconds, as this is when we take model dumps in our simulations. We then applied bolometric corrections to these absolute bolometric magnitudes over approximately seven pulsation cycles at full amplitude. We then phased these regularly spaced absolute bolometric magnitudes in a manner similar to the processing of the observed light curves. This produces a number of points from the seven different cycles of pulsation all overlaid on the same cycle of pulsation. The vertical spread of light curve points indicates the extent of variation of the light curve over the selected cycles. Figures~\ref{fig:LC-T6300_v120}-\ref{fig:LC-T6900_v125} show these synthetic light curves as compared to observational light curves of selected RR Lyrae variables from M3 listed in Table~\ref{table:M3-variables}.

The variables in Table~\ref{table:M3-variables} were selected from 89 observed light curves provided by \cite{Cacciari-2005} of both RRab and RRc variables and phased. The phased light curves where then duplicated for a second cycle for easier visual comparison with the model light curves. These variables were selected by comparing the individual model light curves directly to all the observed variable light curves and selecting the best match as determined from a Bayesian analysis conducted by Michael Gruberbauer considering only the light curve shapes and no model parameters (see \cite{Gruberbauer-2012} for a discussion of this technique with application to fitting solar oscillation modes and \cite{Gregory-2005} for a general introduction to Bayesian analysis). To accomplish the fits only two free parameters were allowed, the distance modulus, and a phase shift, and the parameter space was sampled using nested sampling \citep[e.g.,][]{Feroz-2009}. All variables in Table~\ref{table:M3-variables} were chosen as best matches to the models except v120, which was chosen to compare with a fit by \cite{Marconi-2007} to the same observed light curve with their 1D convective models. \citeauthor{Marconi-2007} adjusted several free parameters (luminosity, mass, effective temperature, and mixing length parameter) in an attempt to match the observed light curve. In comparing their figure~17 to our figure~\ref{fig:LC-T6300_v120}, one can see that our approach to modeling convection produces a significantly better reproduction of the observed light curve for v120. The 1D calculations have difficulty getting the rising slope correctly, and when they do they do not provide a good match to the falling light curve. We have not adjusted any of the free parameters (luminosity, mass or effective temperature) and merely chose representative values of the variable stars in M3 as a whole. While covering this parameter space would be desirable, the computational requirements of our method together with time constraints place this beyond the scope of the current work. Included in these figures are the magnitude differences between the averaged observed light curves and the simulated light curves. The horizontal lines denote differences of 0, $\pm 0.05$, and $\pm 0.08$ magnitudes. For reference, \cite{Marconi-2007} used $\pm 0.05$ and $\pm 0.08$ magnitudes to denote the extrema of their agreement with the observed light curves, although their agreement with observed light curves falls far outside this range for cooler models. Interestingly, the light curves for the cooler 2D models, in which convection is very important, match the observed light curves much more closely than do those of the hotter 2D models.

Particularly good fits are achieved for the cooler models, 6300~K and 6400~K (figures~\ref{fig:LC-T6300_v19} and \ref{fig:LC-T6400_v48} respectively). Hotter models are reasonably good fits given that we have not tuned any of the model parameters to match the observed light curves. There are some features of the light curves that are not matched as well as one might like. In particular model T6700 fit to variable v92 (figure~\ref{fig:LC-T6700_v92}) does not reproduce the dip at minimum light, or get the slope during descending light quite correct. However, adjusting model parameters may be able to account for at least some of the discrepancy. In figure~\ref{fig:LC-T6900_v125} we have compared the light curve of our 6900~K first overtone model with that of v125 and find fairly good agreement, although the model light curve shows a slightly less sinusoidal shape than the observed light curve.

\begin{figure}
\center
\plotone{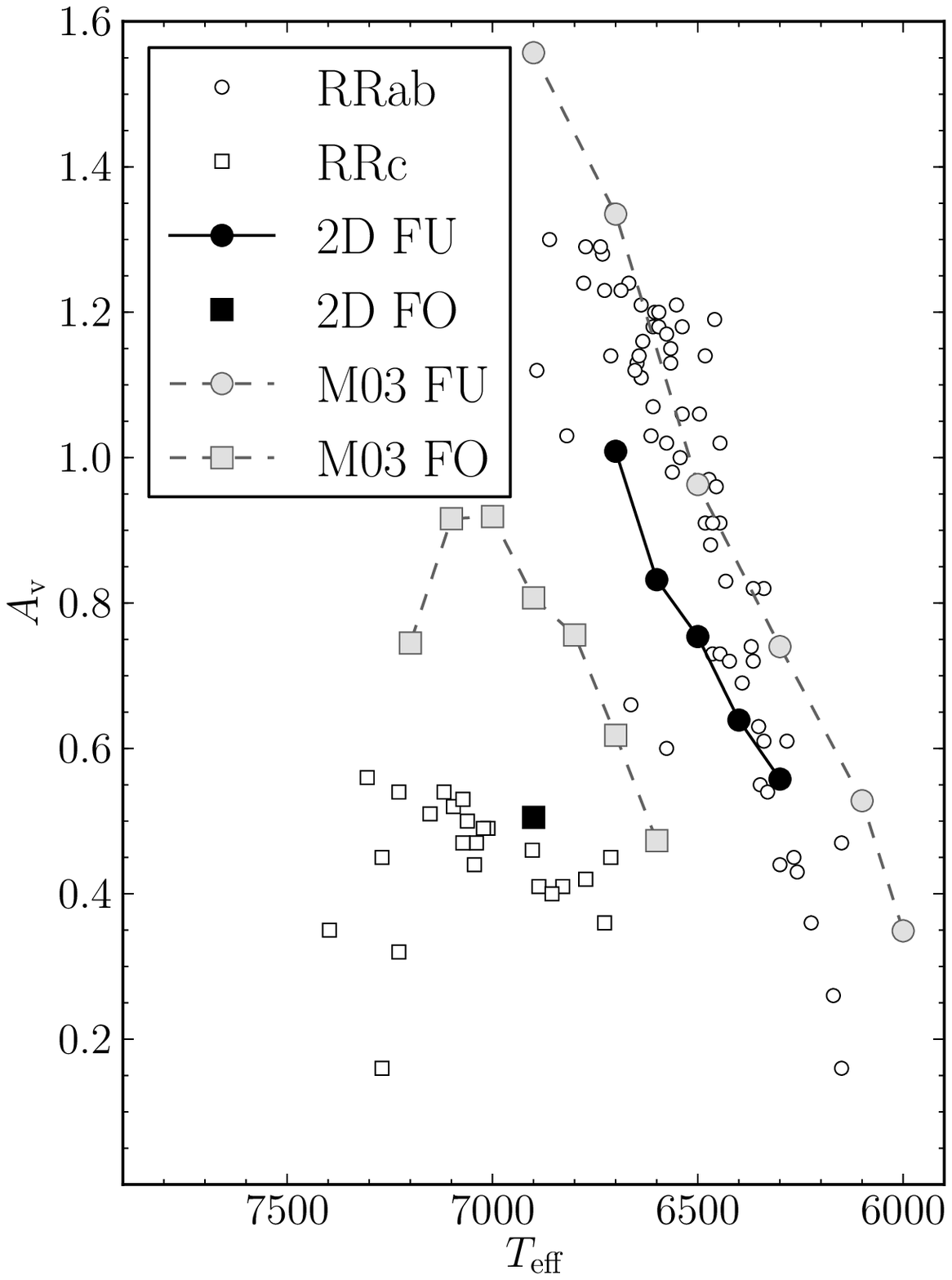}
\caption{Visual pulsation amplitude, $A_V$ across the RR Lyrae instability strip. Circles and squares denote data points for the fundamental and first overtone modes respectively. Filled symbols correspond to models, while open symbols correspond to observations. Black symbols with solid lines are for our 2D convective calculations, and grey symbols with dashed lines are for 1D convective models of \cite{Marconi-2003}.}
\label{fig:AV-v-temp}
\end{figure}

In addition we can compare pulsation amplitude as a function of effective temperature with that derived from observations. Figure~\ref{fig:AV-v-temp} shows the observed visual pulsation amplitudes for fundamental pulsators (open circles) and first overtone pulsators (open squares) in M3 as compared to our model results for fundamental (black filled circles and solid lines) and first overtone modes (filled black square). This figure also contains results from 1D convective models by \cite{Marconi-2003} for the fundamental (filled grey circles with dashed line), and first overtone modes (filled grey squares dashed line). The 1D convective models were for a mass of 0.65~M$_\odot$, Z=0.001, and $\log_{10}(L/L_\odot)=1.61$, while our models were for a mass of 0.7~M$_\odot$, a metallicity of Z=0.0005, and a luminosity of $\log_{10}(L/L_\odot)=1.7$. At least some of the differences between our visual amplitudes and those of \citeauthor{Marconi-2003} result from the differences in mass and luminosity, and to a lesser extent metallicity. Figure 10 of \cite{Bono-1997b} shows that as the mass is increased in their 1D convective models, the amplitude of the first overtone mode increases for a given effective temperature. This effect is less pronounced for the fundamental mode. \citeauthor{Bono-1997b}'s figures~9 and 10 show that as the luminosity is increased the pulsation amplitude of the fundamental mode increases. In the case of the first overtone pulsators however, as the luminosity is increased the pulsation amplitude decreases and the characteristic ``bell'' shape flattens. Based on the trends mentioned for the pulsation amplitude from \cite{Bono-1997b} one might expect that 1D models for the same luminosity and mass as our models would have a flatter ``bell'' shape for the first overtone mode models, and perhaps a slightly larger amplitude for the fundamental mode. Both our calculations and the 1D calculations of \citeauthor{Marconi-2003} produce similar slopes, for the relation between the visual amplitude of the fundamental mode as a function of effective temperature. However, to make direct comparison between the 2D and 1D calculations they would need to be calculated for the same set of model parameters.

To complicate things further, in addition to the mass, luminosity and metallicity, the helium abundance can also play a role in determining the amplitude of the pulsation, though less substantially than the mass or luminosity \cite[see][figure 9]{Bono-1997b}. Finally the pulsation amplitude in 1D models is also influenced by the choice for the mixing length parameter \citep[see][figure~14]{Marconi-2003}, the correct choice of which may vary for different models based on many different variables such as the pulsation mode and the effective temperature. 

In our models the eddy viscosity parameter ($C$ in Equation~(\ref{eq:smag-edd}) ) is a free parameter. We have preformed test calculations where $C$ has been increased and decreased by a factor of two, thus modifying the eddy viscosity, $\mu_t$, by a factor of four. Increasing $C$ by a factor of two decreases both the visual amplitude and growth rate of the test model by 10\%. Conversely the decrease in the eddy viscosity parameter produced an increase in both the growth rate and visual amplitude by about 10\%. We note that a reasonably small change in effective temperature (100~K) can measurably change the visual amplitude (0.2~mags as seen in figure~\ref{fig:AV-v-temp}) and that measurements of effective temperature have uncertainties of this order.

Given these considerations, we would argue that we have done well in matching the observed visual amplitude dependence on the effective temperature, and perhaps some adjustment of our free parameters of mass and luminosity and composition would give an even better match in individual cases.

The reasonable agreement between the observed and model light curves as well as the good agreement between the observed and model visual amplitude dependence on the effective temperature indicate that our use of 2D convective radial pulsation models using large eddy simulations produce light curves that match those of real RR Lyrae stars reasonably well. One should note that this does not necessarily mean that we have a great model for turbulent convection. We believe it is likely that the time dependent behaviour of the convective flux shown in figures~\ref{fig:con-time-dep-6300} and \ref{fig:con-time-dep-6700} is a key component in determining the effect of convection on pulsation, and this behaviour may result with an approach to convection which need not be correct in every detail.

\section{Model Structure with Convection Only}
\label{sec:Con-only}
\begin{figure}[tb!]
\center
\plotone{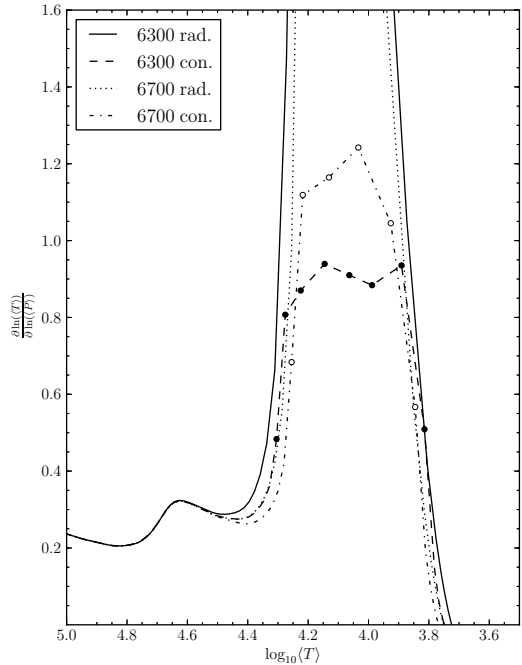}
\caption{Logarithmic gradient of the horizontally averaged temperature with respect to the horizontally averaged pressure versus the horizontally averaged temperature. Four models are represented: $T_{\rm eff} = 6300$~K, radiation only (solid), $T_{\rm eff} = 6700$~K, radiation only (dotted), $T_{\rm eff} = 6300$~K with convection (dash), and $T_{\rm eff} = 6700$~K with convection (dash-dot). The locations of some discrete points have been added for clarity.}
\label{fig:gradient}
\end{figure}

While the emphasis of this research is on the interaction between convection and stellar pulsation, we can also compute the horizontally averaged structure of a model which is not pulsating. This structure will still retain some time dependence because it is a hydrodynamic calculation, but the variation is relatively slight over the 40~day time period between the time when convection had grown large enough to begin affecting the structure to when pulsation, as indicated by periodic variation of the stellar radius, began to affect the structure. We have performed such calculations for models with effective temperatures of 6300~K and 6700~K. The results are presented in figure~\ref{fig:gradient} in the form of the logarithmic derivative of the horizontally averaged temperature with respect to the horizontally averaged pressure versus the horizontally averaged temperature. We have included the structure of both the initial radiation only model and a model when the changes in the structure due to convection appear to be rather stable. The decrease in the gradient in the hydrogen and first helium ionization regions when convection is included is clear. Furthermore, there are some differences in the gradient from that of the radiation only models up to temperatures of about $3 \times 10^4$~K. The effects are larger for the cooler model, as would be expected.

Radiative models of RR Lyrae and Cepheid variables have very steep temperature gradients in the hydrogen ionization zone because the high opacity requires these steep gradients to deliver the flux. As seen in Figure 15, convection decreases the horizontally averaged temperature gradient and spreads out the hydrogen ionization region. In any given angular zone, the radial transition through a temperature of about $1\times 10^4$~K remains relatively abrupt, but the transition occurs in different radial zones for different angular zones so that the horizontally averaged gradient is smoothed out.

\section{SUMMARY}

While the detailed aspects of turbulent convection may not be necessarily correct because of the horizontal extent, the resolution of the problem, the use of a specific subgrid scale model for convective effects on scales we do not resolve, and most importantly because of the 2D nature of the computations, we believe these calculations represent a reasonable compromise between numerical resolution and the ability to perform the calculations for the length of time required on present day computers. The goal is not so much to present a detailed model of convection as to produce a reasonable model of the interaction of convection and large amplitude radial pulsation. A key ingredient appears to be that the pulsation phase dependence of convection carry the right amount of flux in the right place at the right phase. This phase dependence appears to be independent of the strength of convection in our simulations. An exception may be the first overtone model, T6900, in which the convective flux is quite small and relatively noisy.  The amount of flux carried can depend on the physical extent of the simulation, the parameters of the subgrid-scale turbulence model, and the structure of the initial model, while the location of convection is dictated by the steepness of the temperature gradient which depends on the equation of state and opacities.

A comparison of 2D full amplitude light curves to observed light curves in M~3 indicates that 2D convective simulations agree as well as 1D time dependent mixing length convective models at higher effective temperatures, while they appear to be better at lower effective temperatures near (but not at) the red edge. We note that our 2D models have not produced a red edge because the convection zone penetrates sufficiently deep to change the potential energy of the stellar model, thus interfering with the growth of the peak kinetic energy as a pulsational stability discriminant.

The new 2D calculations presented here have been made possible by two key advances - faster computers in larger clusters and the use of a ``Lagrangian'' coordinate in the radial direction made possible by introducing a radial grid velocity that maintains a constant net mass in a given radial shell while allowing natural convective flow through the inner and outer boundaries of the radial shells, as described in Paper I.

In a future paper the results for 3D calculations will be compared to these 2D results, including how the 2D and 3D convective flow patterns differ and how they affect the radial pulsation.

\acknowledgments
These calculations were performed in part on high performance computer clusters provided by ACEnet. ACEnet is funded by the Canada Foundation for innovation and provincial funding agencies, including the Nova Scotia Research Innovation Trust. CMG was supported by Canada Research Chair funds provided to RGD and through an ACEnet Research Fellowship. CMG received partial financial support during writing from a Consolidated STFC grant (ST/J001627/1). Thanks go to Michael Gruberbauer for his Bayesian analysis.

\bibliographystyle{apj}

\end{document}